\documentclass[twocolumn,A4]{article}
\usepackage[dvips]{graphics}

\begin{document}

\onecolumn

\begin{center}
{\bf{\Large Persistent current in one-dimensional non-superconducting
mesoscopic rings: effects of single hopping impurity, in-plane electric
field and foreign atoms}} \\
~\\
Santanu K. Maiti$^{\dag,\ddag,*}$ \\
~\\
{\em $^{\dag}$Theoretical Condensed Matter Physics Division,
Saha Institute of Nuclear Physics, \\
1/AF, Bidhannagar, Kolkata-700 064, India \\
$^{\ddag}$Department of Physics, Narasinha Dutt College,
129, Belilious Road, Howrah-711 101, India} \\
~\\
{\bf Abstract}
\end{center}
Persistent current in one-dimensional non-superconducting mesoscopic rings
threaded by a slowly varying magnetic flux $\phi$ is studied based on the 
tight-binding model. The behavior of the persistent current is discussed in
three aspects: (a) single hopping impurity, (b) in-plane electric field
and (c) in presence of some foreign atoms.
\vskip 1cm
\begin{flushleft}
\end{flushleft}
\vskip 5in
\begin{flushleft}
{\bf $^*$Corresponding Author}: Santanu K. Maiti

~Electronic mail: santanu.maiti@saha.ac.in
\end{flushleft}

\newpage

\section{Introduction}

Advances in nanoscience and technology have made it possible to fabricate
devices in sub-micrometer scale and the transport in such devices gives
several novel and interesting new phenomena. For such mesoscopic systems, 
at sufficiently low temperatures, the semi classical theory of electronic 
transport breaks down. Here, two new important features occur. First, the 
system provides discrete electronic energy levels. Second, the motion of 
an electron is coherent in the sense that once the electron can propagate 
across the whole system without inelastic scattering, its wave function will 
maintain a definite phase. The electron will thus be able to exhibit variety 
of novel and interesting quantum interference phenomena. In what follows, we 
shall concentrate to one of the most striking evidences, called the 
persistent current in normal conducting loops. Starting from 1983, people 
are studying extensively the phenomenon of persistent current in mesoscopic 
one-channel rings, multi-channel cylinders and some twisted geometries both 
theoretically$^{1-17}$ as well as experimentally.$^{18-22}$ Several aspects 
of persistent current those were observed experimentally can be explained 
by theoretical arguments, but till now there are lot of controversies 
between them.

One such promising discrepancy is the amplitude of the measured currents which 
is an order of magnitude larger than the theoretically estimated value. It was 
believed that the electron-electron correlation ($U$) has a significant role to 
enhance the current amplitude in dirty systems. Some perturbative calculations
have also been used to solve this problem and predicted some intuitive results,
but no such clear explanations have yet been found. Most of the theoretical 
works those were performed to explain the combined effects of electron-electron
correlation and impurity on persistent current are basically based on the 
tight-binding model with nearest-neighbor hopping (NNH) integral. This simple
NNH model cannot explain the desired current amplitudes those are observed in 
the experiments. It is quite reasonable and also physical to take higher order 
hopping integrals in addition to the nearest-neighbor hopping, and in the 
theoretical papers$^{23-25}$ we have shown that higher order hopping 
integrals have significant contribution to enhance the current amplitude in 
the presence of impurity. 

Another one important controversy comes from the determination of the sign of 
low-field currents. In an experiment on $10^7$ isolated mesoscopic Cu rings,
Levy {\em et al.}$^{20}$ have measured diamagnetic response of persistent
currents at very low fields, while Chandrasekhar {\em et al.}$^{18}$ 
have determined $\phi_0$ periodic currents in Ag rings with paramagnetic
response for these fields. Theoretically Cheung {\em et al.}$^3$ have
predicted that the direction of persistent current is random depending on the 
total number of electrons, $N_e$, in the system and the specific realization 
of the randomness. Both diamagnetic and paramagnetic responses have been 
observed theoretically in mesoscopic Hubbard ring by Yu and Fowler.$^{26}$ 
They have shown that the rings with odd $N_e$ exhibit paramagnetic response, 
while those with even $N_e$ give diamagnetic response in the limit $\phi 
\rightarrow 0$. In a recent experiment Jariwala {\em et al.}$^{27}$ have
got diamagnetic persistent currents with  both integer and half-integer 
flux-quantum periodicities in an array of $30$-diffusive mesoscopic gold rings.
The diamagnetic sign of the currents in the vicinity of zero magnetic field 
were also found in an experiment$^{21}$ on $10^5$ disconnected Ag ring. 
The sign is a priori not consistent with the theoretical predictions for the 
average of persistent current. In the theoretical paper$^{28}$ we have 
introduced in detail about the sign of the low-field ($\phi\rightarrow 0$) 
currents both for one- and multi-channel mesoscopic loops to understand the 
controversies between these different predicted results. 

Even though different theoretical models were used to explain several
experimental results but a clear understanding of the experiments is still 
lacking. Here we study the behavior of persistent currents in one-dimensional 
mesoscopic rings and focus our attention to the effects of the single hopping 
impurity, the electric field in the plane of the ring and the existence of 
foreign atoms in different lattice sites on these currents. Depending on the 
values of the single hopping impurity we get three different regimes, 
ballistic, metallic and insulating phases respectively. The effect of the 
in-plane electric field on the current is also quite interesting. It is 
observed that the electric field shifts the electronic spectrum and damps 
the amplitude of persistent current. This behavior can be used to control 
the energy spectra and the amplitude of persistent currents {\em externally}. 
Now in the study of the effect of foreign atoms on persistent currents, we 
assume that only in these foreign atoms electron-electron correlation exists 
(here we neglect the electron correlation in parent atoms) and focus several 
interesting new results. 

Our scheme of this paper is as follow. In Section $2$, we study persistent 
current in one-dimensional non-interacting mesoscopic rings in the presence 
of single hopping impurity. Section $3$ provides the effect of in-plane 
electric field on persistent current in one-dimensional non-interacting rings. 
In Section $4$, we discuss the effect of e-e correlation, exists only in the
foreign atoms, on persistent current. Finally, we summarize our results in 
Section $5$. 

\section{One-dimensional mesoscopic ring with single hopping impurity}

\begin{figure}[ht]
{\centering \resizebox*{5.5cm}{3.25cm}{\includegraphics{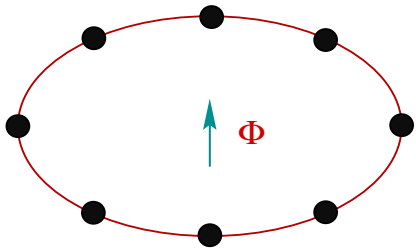}}\par}
\caption{One-dimensional normal metal ring threaded by a magnetic flux $\phi$.
Filled circles correspond to the position of the atomic sites.}
\label{ring}
\end{figure}

The system under consideration is a one-dimensional non-interacting 
mesoscopic ring (Fig.~\ref{ring}) with single hopping impurity. Such a ring 
with $N$ atomic
sites is modeled by a single-band tight-binding Hamiltonian within a
non-interacting picture, which can be written as,
\begin{eqnarray}
H&=&\sum_{i=1}^N \epsilon_i c_i^{\dagger}c_i + v\sum_{i=1}^{N-1}
\left(e^{i\theta}
c_i^{\dagger}c_{i+1}+e^{-i\theta} c_{i+1}^{\dagger}c_i\right) \nonumber \\
& & + v(1-\rho)\left(e^{i\theta}
c_N^{\dagger}c_1+e^{-i\theta} c_1^{\dagger}c_N\right)
\label{hamil1}
\end{eqnarray}
where $\epsilon_i$'s are the on-site energies, $c_i^{\dagger}$ ($c_i$) is the
creation (annihilation) operator of an electron at site $i$, $v$ gives the 
hopping strength between two nearest-neighbor sites and $\theta=2\pi\phi/N$, 
the phase factor due to the flux $\phi$ threaded by the ring. The single 
hopping impurity in this tight-binding Hamiltonian is inserted between the 
sites $1$ and $N$, and the hopping strength can be controlled by the parameter 
$\rho$. Depending on the value of $\rho$, three possible cases appear. 
Case I : $\rho=0$, the system is free from any impurity. This system shows
purely ballistic nature. Case II : $\rho=1$, here electrons cannot hop between 
the sites $1$ and $N$ and accordingly, system goes to the insulating phase. 
Thus for such a case no current will appear. Case III : $0<\rho<1$, here 
ring is treated as a single hopping impurity system or metallic system. 
For all these non-interacting cases, the spin of the electrons does not give 
any new qualitative behavior in the persistent currents and accordingly, in 
this section we ignore the spin dependent term. Throughout this article we 
take $v=-1$ and use the units where $c=e=h=1$. 

At absolute zero temperature, persistent current in the ring threaded by
a magnetic flux $\phi$ is determined from the expression,
\begin{equation}
I(\phi) = - \frac{\partial{E_{0}(\phi)}}{\partial{\phi}}
\label{current}
\end{equation}
where $E_{0}(\phi)$ is the ground state energy.

\begin{figure}[ht]
{\centering \resizebox*{8.5cm}{11cm}{\includegraphics{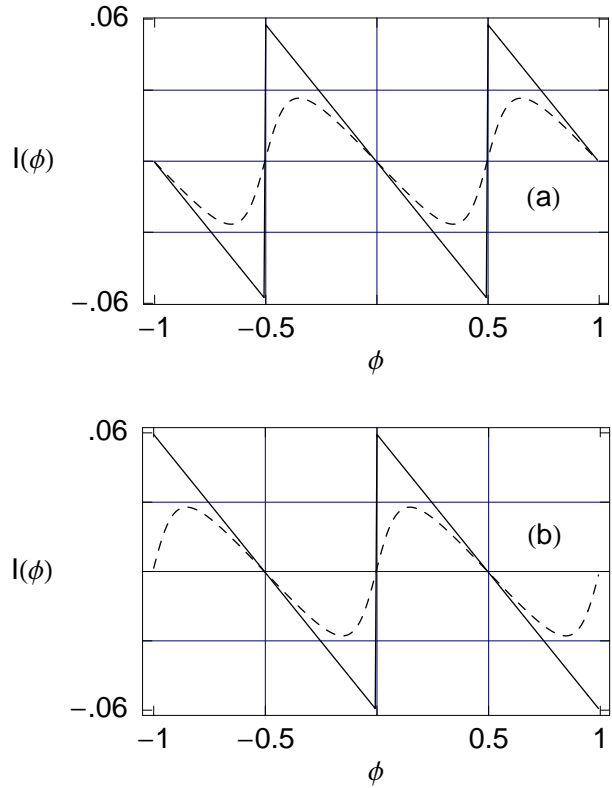}}\par}
\caption{Current-flux characteristics for some one-dimensional non-interacting
mesoscopic rings, where (a) $N_e=75$ and (b) $N_e=80$. Here we take the ring 
size $N=200$. The solid and the dashed curves are respectively for the rings
with $\rho=0$ and $0.5$.}
\label{hopcurr}
\end{figure}

In Fig.~\ref{hopcurr}, we plot the current-flux characteristics for
some typical one-dimensional non-interacting mesoscopic rings taking the 
ring size $N=200$. Figures~\ref{hopcurr}(a) and (b) correspond to the rings 
with $N_e=75$ (odd $N_e$) and $N_e=80$ (even $N_e$) respectively. The solid 
curves 
represent the persistent currents for the rings with $\rho=0$, while the 
dashed curves correspond to the currents for the rings with $\rho=0.5$. In 
the absence of any impurity, the current shows saw-tooth like nature (see solid 
curves) as a function of the flux $\phi$. It is observed from the solid curves 
that, at half-integer or integer multiples of the flux-quantum $\phi_0$, the
current has a sharp transition. This is due of the existence of the degenerate 
energy eigenstates at these respective field points. But as long as the 
impurities are introduced, all these degeneracies move out and the current 
varies continuously with $\phi$ and achieves much reduced value (see dashed 
lines). This reduction of the current amplitude is due to the localization 
effect$^{29,30}$ of the energy eigenstates caused by the presence of 
impurity in the rings. Both for the perfect and dirty rings, the currents 
exhibit $\phi_0$ periodicity. 

For $\rho=1$, the system goes to the insulating phase and no current will be 
available. 

\section{One-dimensional mesoscopic ring with in-plane electric field}

In this section we describe the effect of in-plane electric field on 
persistent current of a one-dimensional non-interacting mesoscopic ring 
\begin{figure}[ht]
{\centering \resizebox*{6.25cm}{3.25cm}{\includegraphics{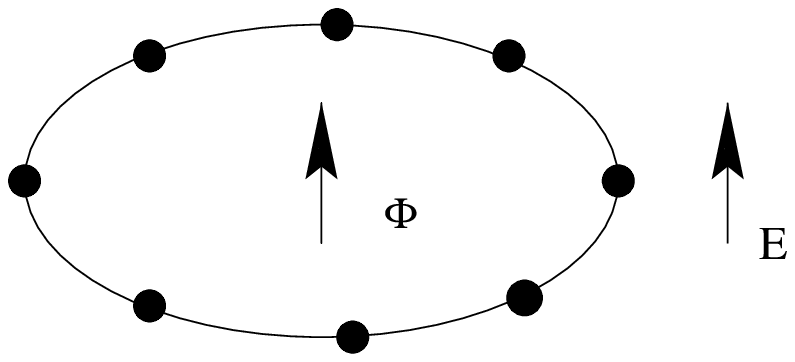}}\par}
\caption{Schematic view of a one-dimensional ring threaded by a magnetic flux 
$\phi$ in the presence of in-plane electric field ($E$). Filled circles 
correspond to the position of the atomic sites.}
\label{ring1}
\end{figure}
(Fig.~\ref{ring1}) within the tight-binding framework. For a $N$-site ring, 
the single-band tight-binding Hamiltonian is written as,
\begin{equation}
H=\sum_{i=1}^N \epsilon_i c_i^{\dagger}c_i + v\sum_{i=1}^N
\left(e^{i\theta} c_i^{\dagger}c_{i+1}+e^{-i\theta} c_{i+1}^{\dagger}c_i\right) 
\label{hamil2}
\end{equation}
where the symbols carry their usual meaning as in Eq.~\ref{hamil1}. Due to the
in-plane electric filed, site energy gets modified and is expressed through 
the relation,$^{31}$
\begin{eqnarray}
\epsilon_i &=& \left(e E a N/2 \pi\right)\cos\left[2 \pi (i-1)/N\right]
\nonumber \\
 &=& \left(ev\right)\left(E^{\star}N/2\pi\right)\cos\left[2 \pi (i-1)/N\right]
\end{eqnarray}
where $E$ is the electric field and $a$ is the lattice spacing. 
Here we define the dimensionless electric field $E^{\star}=Ea/v$.

\begin{figure}[ht]
{\centering \resizebox*{8.5cm}{11cm}{\includegraphics{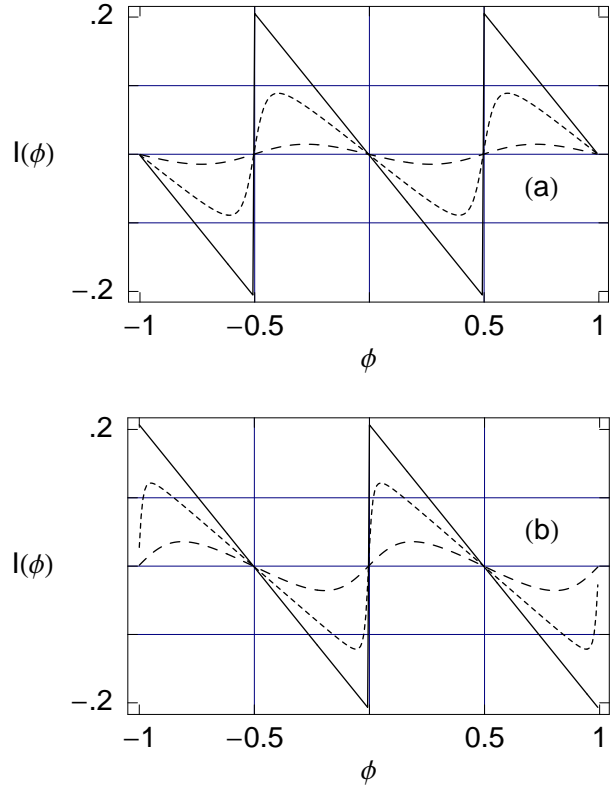}}\par}
\caption{Current-flux characteristics for one-dimensional non-interacting
mesoscopic rings in the presence of in-plane electric field, where 
(a) $N_e=27$ and (b) $N_e=32$. Here we set the ring size $N=60$ and the
lattice spacing $a=1$. The solid, dotted and dashed curves correspond to 
the electric field $E^{\star}=0$, $0.18$ and $0.2$ respectively.}
\label{elecurr}
\end{figure}

Figure~\ref{elecurr} shows the current-flux characteristics for some 
one-dimensional non-interacting rings ($N=60$) in the presence of 
in-plane electric field. The behavior of the persistent currents for 
the rings with odd number of electrons ($N_e=27$) are shown in 
Fig.~\ref{elecurr}(a), while for the rings with even $N_e$ ($N_e=32$) the 
results are shown in Fig.~\ref{elecurr}(b). In the absence of any electric 
field,
the currents exhibit saw-tooth like behavior (see solid curves) with sharp 
transitions at half-integer (for odd $N_e$) and integer (for even $N_e$) 
multiples of the elementary flux quantum $\phi_0$, similar to that as given 
by the solid curves in Fig.~\ref{hopcurr}. On the other hand, in the 
presence of in-plane electric field, the saw-tooth like behavior of the 
currents disappears and the currents vary continuously with $\phi$ as shown 
by the dotted ($E^{\star}=0.18$) and the dashed ($E^{\star}=0.2$) curves. 
Our numerical results predict that, the current amplitude gets reduced with 
the increase of the electric field which is really an interesting
one. It is also examined that the current amplitude decays exponentially 
with this electric field (not shown here in the figure). Thus we can 
emphasize that, the behavior of the persistent current in one-dimensional 
non-interacting rings with an in-plane electric field is quite similar to 
that of one-dimensional non-interacting rings in the presence of impurity, 
but the significant feature is that in the previous systems i.e., the rings 
with in-plane electric field, one can control the current amplitude 
{\em externally} by tuning the electric field which provides a key 
idea for fabrication of efficient nano-scale devices. 

\section{One-dimensional mesoscopic ring with foreign atoms in various 
lattice sites}

This section demonstrates persistent currents in one-dimensional mesoscopic 
ring with foreign atoms in different lattice sites. The speciality of the 
foreign atoms is that, only in these atoms the electron-electron correlation 
exists, while the parent atoms are free from any such interaction. Actually 
such systems can be observed where one dopes some foreign atoms in the parent 
system and we also notice such systems in reality. In the previous two 
sections, we have studied the persistent currents in normal conducting rings 
within the 
framework of one-electron picture, but in the presence of e-e correlation, 
we have to consider the many-body Hamiltonian to describe our model. The 
tight-binding model Hamiltonian for an interacting ring with $N$ atomic 
sites is expressed in this form,
\begin{eqnarray}
H &=& \sum_{i=1,\sigma}^N\epsilon_{i,\sigma}c_{i,\sigma}^{\dagger}c_{i,\sigma}
+ v \sum_{i=1,\sigma}^N \left(e^{i\theta}c_{i,\sigma}^{\dagger}c_{i+1,\sigma}+
e^{-i\theta}c_{i+1,\sigma}^{\dagger}c_{i,\sigma}\right) \nonumber \\
& & + U\sum_{d=1}^{N_d} c_{d,\uparrow}^{\dagger}c_{d,\uparrow}
c_{d,\downarrow}^{\dagger}c_{d,\downarrow}
\label{hamil3}
\end{eqnarray}
where $c_{d,\sigma}^{\dagger}$ ($c_{d,\sigma}$) is the creation 
(annihilation) operator of an electron with spin $\sigma$ ($\uparrow$ or
$\downarrow$) at site $d$ where the foreign atom situates. The parameter 
$U$ corresponds to the strength of the Hubbard 
correlation, exists only in $N_d$ foreign atoms where $N_d \leq N$. 
In these interacting systems, since the dimension of the many-body 
Hamiltonian matrices increase very sharply with $N$ for higher number 
of electrons $N_e$, and also the computational operations are so time 
consuming, we restrict our study only on the systems of smaller $N$ 
and $N_e$. In what follows, we shall describe our results for the few 
cases those are respectively given as: 
(I) ring with two opposite ($\uparrow,\downarrow$) spin electrons, 
(II) ring with three ($\uparrow,\uparrow,\downarrow$) spin electrons, 
(III) ring with four ($\uparrow,\uparrow,\downarrow,\downarrow$) spin 
electrons and (IV) ring with five ($\uparrow,\uparrow,\uparrow,\downarrow,
\downarrow$) spin electrons.

At absolute zero temperature ($T=0$), the persistent current in such 
interacting rings is determined by taking the first order derivative of the 
many-body ground state energy $E_0(\phi)$, which is obtained from the exact 
numerical diagonalization of the tight-binding Hamiltonian Eq.~\ref{hamil3}.

\subsection{Ring with two ($\uparrow,\downarrow$) spin electrons}

\begin{figure}[ht]
{\centering \resizebox*{12cm}{9cm}{\includegraphics{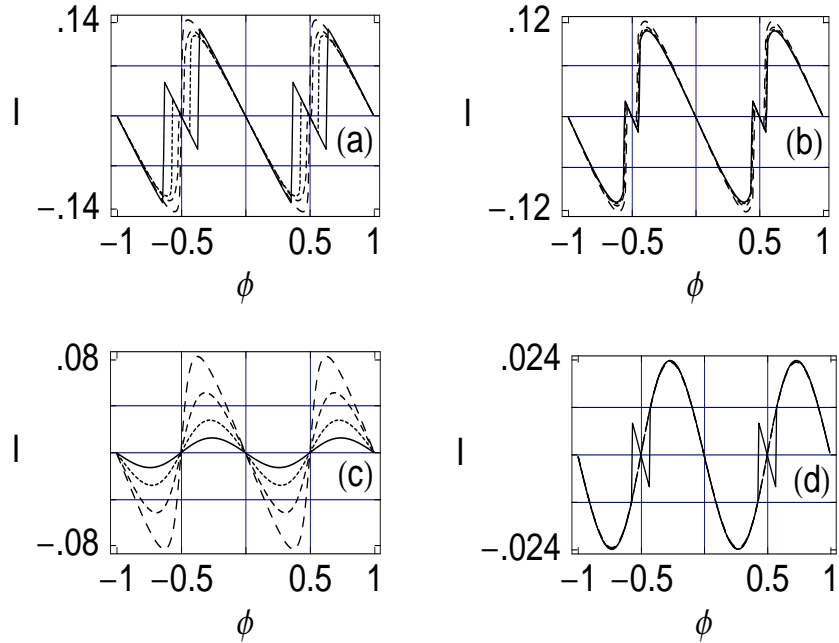}}\par}
\caption{Current-flux characteristics for some interacting rings ($N=20$) with 
two opposite ($\uparrow,\downarrow$) spin electrons. The dashed, dotted, 
small dotted and solid lines in different figures are respectively for: 
(a) $N_d=4$, $8$, $12$, $16$ with $U=2$; (b) $U=2$, $4$, $6$, $8$ with 
$N_d=10$; (c) $N_d=4$, $8$, $12$, $16$ with $U=2$ and $W=1$; (d) $U=2$, $4$, 
$6$, $8$ with $N_d=10$ and $W=1$.}
\label{two}
\end{figure}

To understand the precise dependence of the electron-electron correlation, 
exists in foreign atoms, on persistent current let us first take the simplest 
possible system which is the case of a ring with two opposite spin electrons.
Figure~\ref{two} shows the current-flux characteristics of some 
one-dimensional rings ($N=20$) with two opposite spin electrons.

Figure~\ref{two}(a) gives the results for the rings in the absence of any 
impurity ($W=0$, where $W$ is the strength of the randomness) with Hubbard 
correlation strength $U=2$. The dashed, dotted, small dotted and solid curves 
correspond to the rings with $N_d=4$, $8$, $12$ and $16$ respectively. It is 
observed that, suddenly the direction and the magnitude of the current change 
across $\phi=\pm \phi_0/2$ and a kink-like structure appears in the current. 
With the increase of $N_d$, the length of the kink increases, which is clearly 
visible from the curves plotted in this figure. Though the effective Hubbard 
correlation strength increases with the increment of $N_d$, but yet the 
kinks are situated at the same region and they overlap with each other. 
This feature clearly manifests that the kinks are independent of the 
correlation strength $U$. The appearance of the kink-like 
structures and their independence on the e-e correlation can be explained 
as follows. For the two opposite spin electron system, total spin $S$ has 
two values which are $0$ and $1$. The Hamiltonian of such a system, 
for any $\phi$, can be block diagonalized by proper choices of the basis 
states. This can be achieved by choosing the basis states from the two 
different sub-spaces with $S=0$ and $S=1$. The sub-space spanned by the 
basis set with $S=1$, the block Hamiltonian is free from $U$, and hence 
the corresponding energy eigenvalues and eigenstates are $U$-independent. 
On the other hand, for the other sub-space with $S=0$, all the energy 
eigenstates are $U$-dependent. In the absence of any electron correlation, 
the $U$-independent energy eigenstates situate always above the ground state 
for any value of $\phi$. But for non-zero values of $U$, one of these 
$U$-independent energy levels achieves the ground state energy of the system 
in certain domains of $\phi$ and the length of these domains increases with 
$N_d$ which produce larger kinks. In these regions, we observe the kinks in 
the $I$-$\phi$ curves, and it is obvious that the persistent currents inside 
these kinks are independent of the Hubbard correlation $U$. 

The behavior of the persistent current in perfect rings with two opposite 
spin electrons is quite similar to that of Fig.~\ref{two}(a) if we fix $N_d$ 
instead of the e-e correlation $U$. As illustrative example, in 
Fig.~\ref{two}(b) we plot the current-flux characteristics for some perfect 
rings considering $N_d=10$. The dashed, dotted, small dotted and solid lines 
correspond to the rings with $U=2$, $4$, $6$ and $8$ respectively, where 
all these curves almost overlap with each other. 

The situation becomes much more interesting when we add impurity in these
rings. Here we assume that the impurities, taken randomly from a ``Box" 
distribution function of width $W$, are given only in $N_d$ atomic sites 
where the e-e correlation exists. The characteristic behavior of the 
persistent currents for some disordered
rings ($W=1$) is shown in Fig.~\ref{two}(c), where we set $U=2$. The dashed,
dotted, small dotted and solid curves correspond to the same values of $N_d$ 
as in Fig.~\ref{two}(a). It is known that the repulsive Coulomb interaction
doesn't favor occupancy of the two electrons in a same site and also it
opposes confinement of the electrons due to localization of energy eigenstates.
Hence the mobility of the electrons increases as we introduce the Hubbard 
interaction and the current amplitude gets enhanced. But this enhancement 
ceases to occur after certain value of $U$ due to the ring geometry, and the
persistent current then decreases as we increase $U$ further. This is the 
basic feature of persistent current in any dirty rings in the presence of 
the electron correlation.
The curves in Fig.~\ref{two}(c) show that the currents vary continuously 
with flux $\phi$ without giving any kink and their amplitudes decrease 
gradually with the increase of $N_d$. This is due to the fact that with the
increase of $N_d$, both the effective electron correlation and randomness
increase but since the later effect dominates over the previous one,
the current amplitude decreases gradually. For these cases, the 
$U$-independent energy eigenstates do not contribute to the lowest energy 
in any energy domain and accordingly, no kink appears in the current.

The dependence of the randomness, the total number of foreign atoms and the
electron-electron correlation on the appearance of kinks in the persistent 
current is much more clearly observed from Fig.~\ref{two}(d), where we plot 
the current-flux characteristics for some disordered rings with fixed $N_d$. 
The different curves in this figure correspond to the same values of $U$ as 
in Fig.~\ref{two}(b). Comparing the results of Figs.~\ref{two}(b) and (d), 
we clearly observe that for the dirty rings the kinks disappear for $U=2$, 
$4$ and $6$, while the kink resides only for $U=10$ (solid curve of 
Fig.~\ref{two}(d)). Thus we can emphasize that the electron-electron 
correlation, randomness and $N_d$ have important significance on the 
behavior of persistent current in such small rings. For all these cases the 
persistent currents exhibit $\phi_0$ flux-quantum periodicity.

\subsection{Ring with three ($\uparrow,\uparrow,\downarrow$) spin electrons}

With the above background we now study the behavior of persistent current in 
mesoscopic interacting rings with higher number of electrons $N_e$. Here we 
\begin{figure}[ht]
{\centering \resizebox*{12cm}{9cm}{\includegraphics{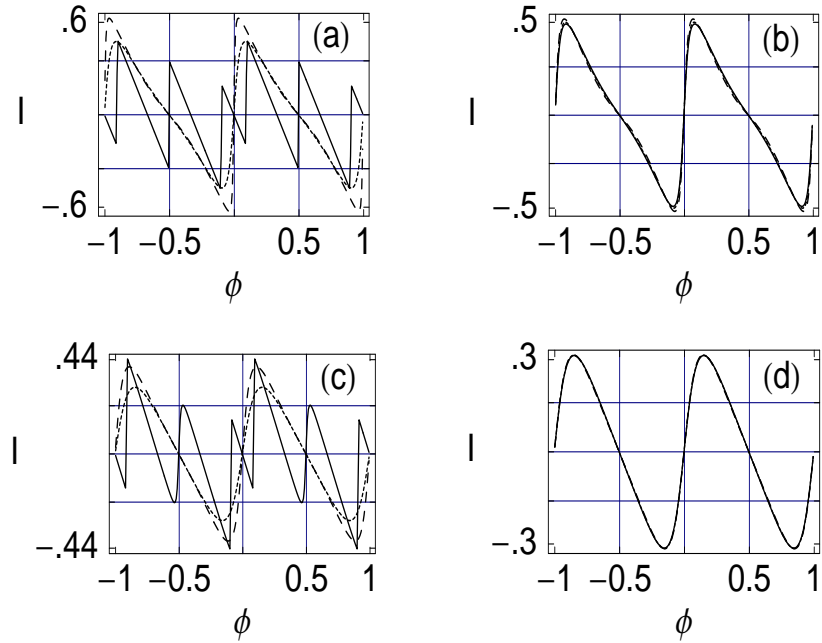}}\par}
\caption{Current-flux characteristics for some interacting rings ($N=10$) with 
three ($\uparrow,\uparrow,\downarrow$) electrons. The dashed, small dotted 
and solid lines in different figures are respectively for: (a) $N_d=4$, $6$, 
$10$ with $U=16$; (b) $U=4$, $6$, $8$ with $N_d=6$; (c) $N_d=4$, $6$, $10$ 
with $U=16$ and $W=1$; (d) $U=4$, $6$, $8$ with $N_d=6$ and $W=1$.}
\label{three}
\end{figure}
consider rings with two up and one down spin electrons as illustrative 
example of three spin electron systems. In Fig.~\ref{three}, we plot the
current-flux characteristics for some of such interacting rings taking the 
ring size $N=10$.

Figure~\ref{three}(a) gives the variation of the persistent currents for the 
rings in the absence of any impurity ($W=0$). Here we set the correlation 
strength
$U=16$. The dashed, small dotted and solid lines are respectively for the
rings with $N_d=4$, $6$ and $10$. The current shows a kink-like structure 
across $\phi=0$ only for the ring with $N_d=10$ (see solid curve). It is 
due the $U$-independent eigenstates like as the two electron systems, as 
explained earlier, and the current inside the kink is independent of the 
strength of the Hubbard correlation $U$. It is also noticed that the kinks 
in the current for such three spin electron systems appear comparatively at 
quite high values than the two spin electron systems.

For the perfect rings described with fixed $N_d$ instead of $U$, no kink
appears in the currents as shown by the curves in Fig.~\ref{three}(b). Here
we put $N_d=6$. The dashed, small dotted and solid lines respectively
corresponds to the current for the rings with $U=4$, $6$ and $8$, where 
all these curves almost coincide with each other. It is observed that for 
all such values of $U$, the $U$-independent energy eigenstates do not 
contribute to the lowest energy in any energy domain and the kinks will 
appear for more higher values of $U$ than those are considered here.    

Now we describe the behavior of the current-flux characteristics for these 
three spin electron systems in the presence of impurity. Figure~\ref{three}(c) 
shows the results for some disordered rings ($W=1$) with same parameters 
as taken in Fig.~\ref{three}(a). The dashed, small dotted and solid curves
correspond to the similar meaning as in Fig.~\ref{three}(a). From this figure
we see that initially the current amplitude decreases, but it again increases 
for higher value of $N_d$ ($N_d=10$, solid curve). This is due to the fact 
that for $N_d=10$, the effective Hubbard interaction dominates over the 
randomness. Here the kink also exists only for $N_d=10$, similar to that 
as observed in Fig.~\ref{three}(a).

The behavior of the persistent current in the presence of impurity for the 
rings specified by the same parameters as in Fig.~\ref{three}(b) is plotted 
in Fig.~\ref{three}(d) and it shows almost similar behavior to that as drawn 
in Fig.~\ref{three}(b). Due to the randomness, the current amplitudes get 
reduced slightly and there is no possibility for the appearance of any 
kink-like structure for these parameter values. For all such rings with 
three spin electrons the current exhibits $\phi_0$ flux-quantum periodicity.

\subsection{Ring with four ($\uparrow,\uparrow,\downarrow,\downarrow$) spin 
electrons}

For a systematic approach, next we focus the behavior of persistent current
in four electron systems and as illustrative example, we consider rings with 
\begin{figure}[ht]
{\centering \resizebox*{12cm}{9cm}{\includegraphics{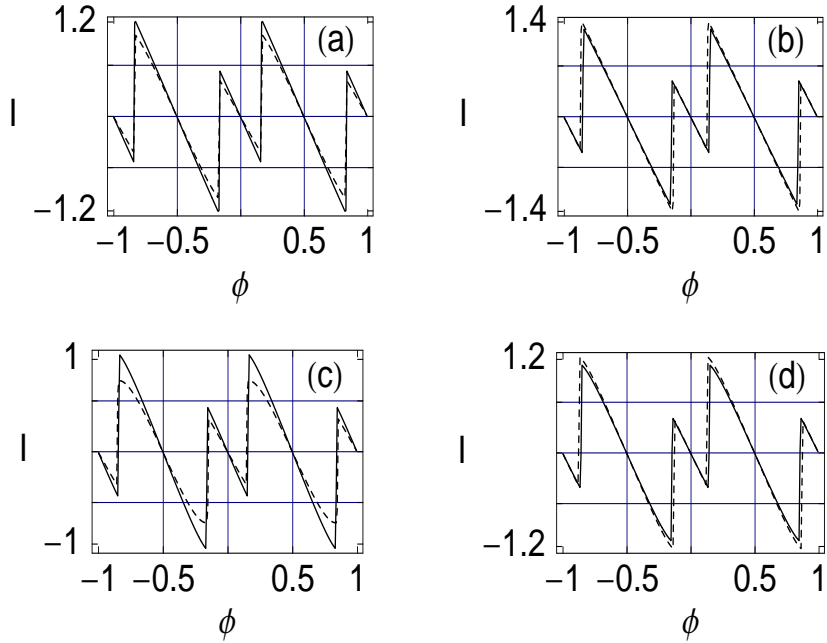}}\par}
\caption{Current-flux characteristics for some interacting rings ($N=8$) with 
four ($\uparrow,\uparrow,\downarrow,\downarrow$) electrons. The dotted and 
solid lines in different figures are respectively for: (a) $N_d=6$, $8$ with 
$U=10$; (b) $U=6$, $8$ with $N_d=8$; (c) $N_d=6$, $8$ with $U=10$ and $W=1$; 
(d) $U=6$, $8$ with $N_d=8$ and $W=1$.}
\label{four}
\end{figure}
two up and two down spin electrons. In Fig.~\ref{four}, we draw the
current-flux characteristics for some of the interacting rings ($N=8$) with 
four spin electrons.

Figure~\ref{four}(a) gives the variation of the persistent currents for some 
perfect rings characterized by constant Hubbard correlation. Here we set 
$U=10$ which is quite high for these rings. The dotted and the solid lines 
in this figure are respectively for $N_d=6$ and $8$. The current shows 
kink-like structure across $\phi=0$ and the appearance of the kinks is due 
to the additional crossing of the ground state energy level with the other
energy levels as we vary the flux $\phi$. Here it is noted that, in the
present case the kinks arise due to the $U$-dependent eigenstates, not from 
the $U$-independent eigenstates as in the earlier cases. Quite similar feature
is also observed for the perfect rings specified with fixed $N_d$ ($N_d=8$),
instead of $U$. The results are plotted in Fig.~\ref{four}(b), where the 
dotted and the solid curves correspond to the currents for the rings 
with $U=6$ and $8$ respectively. 

The current-flux characteristics for some dirty ($W=1$) rings with four spin 
electrons are shown in Figs.~\ref{four}(c) and (d). For both of these two
figures we take the same values of the different parameters those are 
considered respectively in Fig.~\ref{four}(a) and (b). Since the values of
$N_d$ and $U$ are quite large for these rings compared to the system size
$N$, the current shows almost similar variation even in the presence of 
impurity. The persistent current amplitude decreases slightly due to the 
effect of the impurity in the rings and for all these rings the current 
shows only $\phi_0$ flux-quantum periodicity.

We find striking similarity in the behavior of the persistent currents with 
rings containing two opposite spin electrons with the rings containing four 
spin electrons. Hence it becomes apparent that mesoscopic Hubbard rings with 
even number of electrons exhibit similar characteristic features in the 
persistent current.

\subsection{Ring with five ($\uparrow,\uparrow,\uparrow,\downarrow,
\downarrow$) spin electrons}

Lastly, we describe the characteristic behavior of persistent current in five
electron systems and as representative example we take rings with three up 
\begin{figure}[ht]
{\centering \resizebox*{12cm}{9cm}{\includegraphics{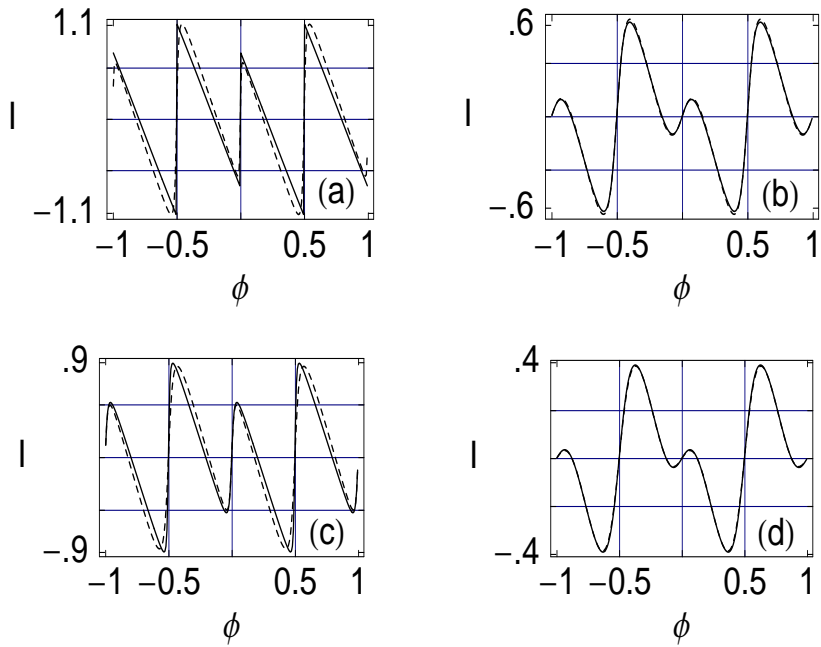}}\par}
\caption{Current-flux characteristics for some interacting rings ($N=7$) with 
five ($\uparrow,\uparrow,\uparrow,\downarrow, \downarrow$) electrons. 
The dotted and solid lines in different figures are respectively for: 
(a) $N_d=5$, $7$ with $U=12$; (b) $U=8$, $12$ with $N_d=6$; (c) $N_d=5$, $7$ 
with $U=12$ and $W=1$; (d) $U=8$, $12$ with $N_d=6$ and $W=1$.}
\label{five}
\end{figure}
and two down spin electrons. Figure~\ref{five} shows the variation of the
persistent currents for some specific rings ($N=7$) with five electrons.

In Fig.~\ref{five}(a), we draw the current-flux characteristic curves for 
some perfect rings described by the constant value of $U$. We set $U=12$ for 
these rings. The dotted and the solid lines are respectively for the rings
with $N_d=5$ and $7$. Though it seems that the strength of Hubbard correlation
is quite high but yet for this value of $U$ no kink-like structure appears in 
the currents and it is noted that the kink will appear for some higher values 
of $U$ (not shown here in this figure). Now the variation of the persistent 
currents for some perfect rings described with fixed $N_d$ is given in 
Fig.~\ref{five}(b), where the dotted and the solid lines are respectively for
$U=8$ and $12$ and they almost overlap with each other. It is observed that
the current amplitude gets reduced, even in these perfect rings, to nearly 
half of the value given in Fig.~\ref{five}(a). This is due to the strong 
repulsive effect caused by the higher value of the effective $U$.

In the presence of impurity, the variation of the persistent currents are 
plotted in Figs.~\ref{five}(c) and (d), where all the parameters have the
same values as considered respectively in Fig.~\ref{five}(a) and (b). 
The feature of these currents are almost similar to that of the perfect 
ring results. Similar to the previous cases, here also the persistent 
currents exhibit $\phi_0$ flux-quantum periodicity.

It is clear from these results that the behavior of the persistent current 
in rings with three spin electrons is quite similar to that of rings with
five spin electrons. Hence it becomes apparent that mesoscopic Hubbard rings 
with odd number of electrons exhibit similar characteristic features in the 
persistent current.

\section{Concluding remarks}

In conclusion, we have investigated the characteristic features of persistent
currents in one-dimensional mesoscopic rings based on the tight-binding model.
Here we have focused our attention to the effects of the single hopping 
impurity, the in-plane electric field and the Hubbard correlation in foreign 
atoms on persistent currents. 

The first part of this article has described the dependence of the persistent 
current on the single hopping impurity, and we have seen that depending on 
the strength of $\rho$, three possible regimes (ballistic ($\rho=0$), 
metallic ($0< \rho < 1$) and insulating ($\rho=1$)) appear. In the presence 
of impurity, the current amplitude gets reduced due to the localization 
effect.

Next we have studied the effect of the in-plane electric field on the
persistent current, and the main motivation for that calculation was to 
observe how one can control the current amplitude {\em externally}, which
provides a key idea for the fabrication of efficient nano-scale devices. 

In the last part, we have described the combined effects of the
electron-electron correlation (exist only in the foreign atoms) and the
randomness on the persistent current. An important finding is the appearance 
of kink-like structures in the current-flux characteristics. Quite 
interestingly we have observed that, in some cases, persistent currents inside 
the kinks are independent of the strength of the interaction $U$. These kinks 
give rise to anomalous Aharonov-Bohm oscillations in the persistent current, 
and recently Keyser {\em et al.}$^{32}$ experimentally observed similar 
anomalous Aharonov-Bohm oscillations in the transport measurements on 
small rings.

\end{document}